\documentclass{camera}
\usepackage{graphicx}

\begin{document}

%%%%%%%%%%%%%%%%%%%%%%%%%%%%%%%%%%%%%%%%%%%%%%%%%%%%%%%%
% The title, only the first letter capitalized; if you want to split it in
% two or more lines, put a \\ macro at each line break
% example: 
%   \title{Title: first line\\ second line}
%
\title{A study of stopping power in nuclear reactions at intermediate energies}

%%%%%%%%%%%%%%%%%%%%%%%%%%%%%%%%%%%%%%%%%%%%%%%%%%%%%%%%
% The author(s), separated by commas; do not put a
% comma before the last author, use instead the \and
% macro which produces a normal ``and'' in the
% caps/small caps context
%
\author{Gr\'egory Lehaut, Dominique Durand \and Olivier Lopez \\for the INDRA Collaboration}

%%%%%%%%%%%%%%%%%%%%%%%%%%%%%%%%%%%%%%%%%%%%%%%%%%%%%%%%
%
\organization{LPC Caen, ENSICAEN, Universit\'e de Caen, CNRS/IN2P3, Caen, France}

\maketitle

\begin{abstract}

We show a systematic experimental study based on INDRA data of the stopping power in central symmetric nuclear 
reactions. Total mass of the systems goes from 80 to 400 nucleons while the incident energy 
range is from 12 AMeV to 100 AMeV. The role of isospin diffusion at 32 and 45 MeV/nucleon 
with $^{124,136}$Xe projectiles on $^{112,124}$Sn targets performed at GANIL is also 
discussed. Results suggest a strong memory of the entrance channel above 20 AMeV/A (nuclear transparency) 
and, as such, constitute valuable tests of the microscopic transport models.

\end{abstract}

%%%%%%%%%%%%%%%%%%%%%%%%%%%%%%%%%%%%%%%%%%%%%%%%%%%%%%%%
% Write the text starting from here and using the usual
% LaTeX commands.
%

The study of transport phenomena in nuclear reactions at intermediate energies is of major importance 
in the understanding of the fundamental properties of nuclear matter~\cite{Dur01}.
The comparison of microscopic transport models~\cite{Aic91,Ono92,Bon94,Cho94,Ohn95,Liu01,Gai05} with 
experimental data can help to improve our knowledge of the basic ingredients of such models: the nuclear 
equation of state and the in-medium nucleon-nucleon cross-section.
In this context, it is necessary to test different models with help of experimental data sets of various system 
sizes and incident energies.

For instance, the FOPI collaboration has studied nuclear stopping in the 200-800 AMeV~\cite{And06} range 
for central Au+Au collisions. In this talk,
we study the low energy "intermediate" range (10 to 100 AMeV) for which a transition between mean 
field effects and two-body dissipation (nucleon-nucleon in-medium collisions) is expected. Symmetric systems with 
total sizes between 80 and 400 nucleons and incident energies from 12 to 100 AMeV measured 
with the INDRA multidetector~\cite{Pou95} at GANIL and at GSI have been measured allowing a 
systematic study of nuclear stopping in central collisions.

\section{Method}

The selection of central collisions is based on a minimum bias selector with a minimal dependence on the kinematical
properties of detected charged particles to avoid auto-correlations. To this end, the total multiplicity of 
charged products has been chosen as an impact parameter selector~\cite{Pha92}: the largest multiplicity being associated to
the most central collisions. The stopping power is estimated by means 
of the isotropy ratio defined by:

\begin{equation}
	E_{iso} = \frac{1}{2} \frac{\sum E_\perp}{\sum E_\parallel}
	\label{equEiso}
\end{equation}

where $E_\perp$ is the transverse kinetic energy and $E_\parallel$ is the longitudinnal kinetic energy, the 
sum running over particles and fragments emitted in the forward part of the center of mass in order to 
minimise the role of detection theresholds. For an isotropic event, $E_{iso}$ is close to 1, and is associated 
to fully stopped events.

\begin{figure}[!htbp]
	\includegraphics[width=1\columnwidth]{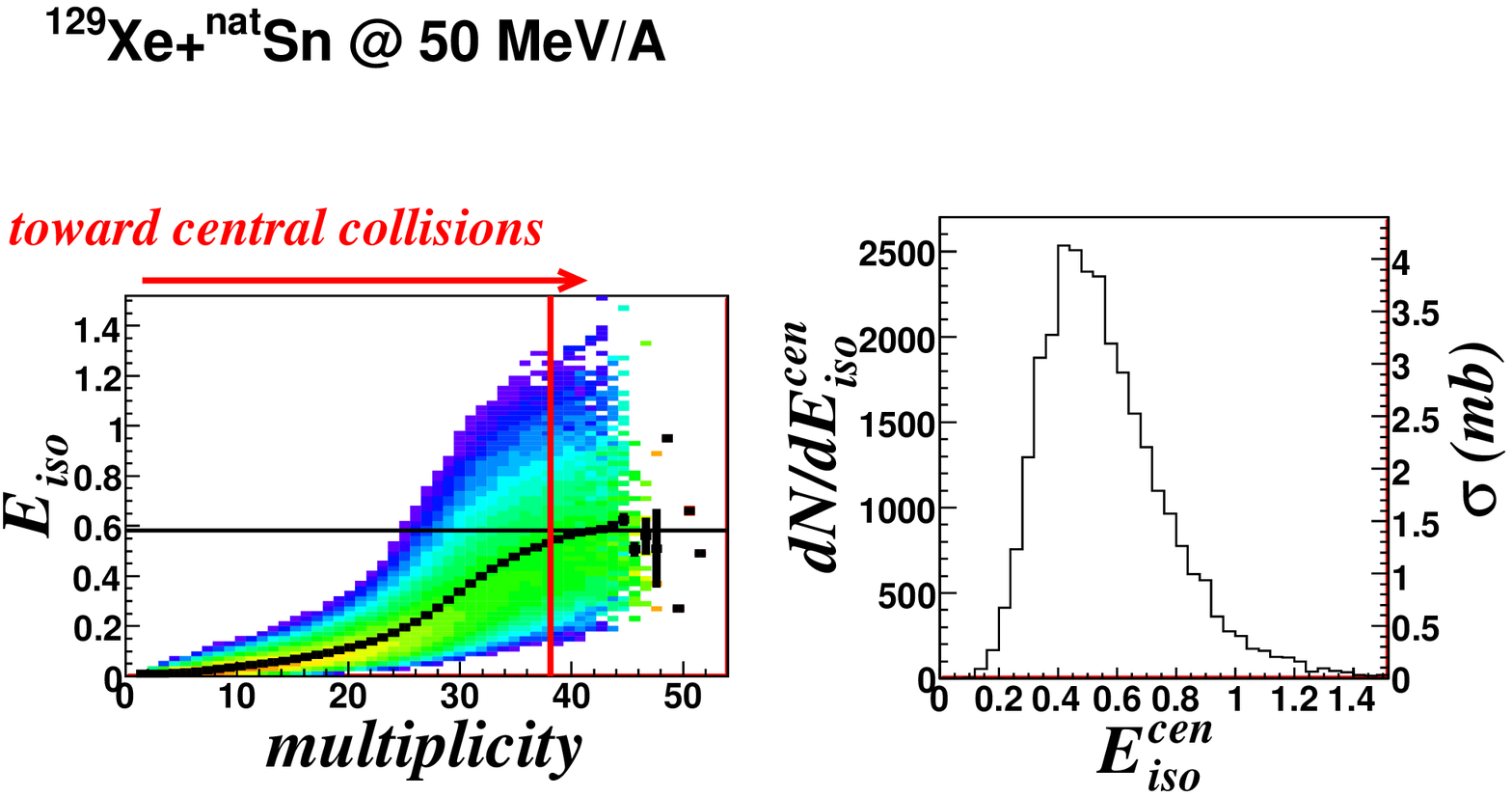}
  \caption{Data for $^{129}$Xe+$^{nat}$Sn collisions at 50 AMeV.
  Left: bi-dimensional plot showing the correlation between the isotropy ratio and the total multiplicity, 
  the red line shows the cut of the multiplicity for central collisions (see text).
  Right: Corresponding $E_{iso}$ distribution for central collisions.}
  \label{fig1} % optional figure label, must be unique
\end{figure}

Figure~\ref{fig1}-left displays the correlation between the multiplicity and $E_{iso}$ for Xe+Sn 
reaction at 50 AMeV. The evolution of the mean value of $E_{iso}$ has been superimposed. 
After an increase of $E_{iso}$ as multiplicity increases, a saturation is observed 
for the largest values of the multiplicity (i.e. most 
central collisions). This leveling off is used as a multiplicity cut to select central 
collisions. Associated cross sections are around 50 mbarns. For such events, 
the $E_{iso}$ distribution is shown on the right part of Fig.~\ref{fig1}. A partial 
stopping of the projectile on the target (i.e. $E_{iso}^{cen}< 1$) is observed. The same procedure has been 
applied for all symmetric reactions measured with INDRA.

\section{Systematic of the stopping power}

\begin{figure}[!htbp]
	\begin{center}
		\includegraphics[width = 0.7\columnwidth]{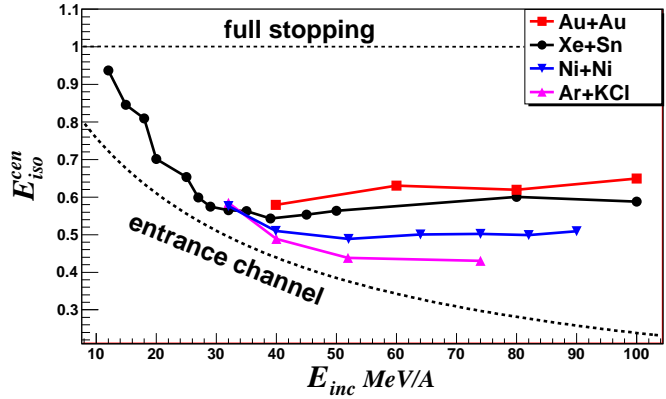}
	\end{center}
	\caption{Evolution of $E_{iso}$ as a function of beam energy. Data are for central 
	collisions and for four different systems indicated in the insert, dashed line $E_{iso}^{cen}$ mean value 
	from initial momentum distribution (see text).}
	\label{fig2}
\end{figure}

Figure~\ref{fig2} shows the evolution of the mean value of $E_{iso}$ for central collisions 
as a function of the beam energy for the different studied systems.

A decrease of $E_{iso}^{cen}$ below 40 AMeV is observed. In such an energy range, the isotropy ratio 
is almost independent of the size of the system. However, as beam energy crosses the Fermi energy  
and for heavier systems, an increase 
of the isotropy ratio depending on the system size is observed: the larger the size, the larger $E_{iso}^{cen}$ although 
for the smallest systems considered in this work, a saturation is present. In any case, low values far from unity are 
observed (i.e. no full stopping). For a sake of comparison, we have calculated the mean value of $E_{iso}$ in the case of 
free nucleons located initially 
in the Fermi sphere of the projectile and of the target:

\begin{displaymath}
<E_{iso}^{init}> = \frac{1}{1+5\frac{p^{2}_{rel}}{p^2_{f}}}
\end{displaymath}

where $p_{rel}$ is the relative momentum between the two Fermi spheres and $p_f$ is the Fermi momentum. The evolution of 
$<E_{iso}^{init}>$ from initial conditions (i.e. entrance channel) as a function of the beam energy 
is show in Figure~\ref{fig2}. The minimum of nuclear stopping power 
is around 40 AMeV, corresponding to a 
smooth transition from a mean-field bahaviour to a two-body dissipation regime.

\section{Isospin effects}

In the framework of the IQMD improved by Liu \emph{et al.} \cite{Liu01}, no effect of the isospin content 
on stopping power for Sn on Sn reactions in the beam energy range from 15 to 150 AMeV has been 
oberved. The INDRA collaboration has measured different isospin content Xe+Sn reactions at 32 and 45 AMeV. Results 
concerning 
the mean value of $E_{iso}^{cen}$ are reported in table~\ref{table1}.

\begin{table}[!htbp]
	\begin{center}
		\begin{tabular}{|c|c|c|c|c|}	
			\hline
			System              & $N/Z$  & label & 32 A.MeV & 45 A.MeV  \\
			\hline
			$^{124}Xe+^{112}Sn$ &  1.27  & PP &  0.54$\pm0.03$    & 0.53$\pm0.03$      \\
			\hline 
			$^{124}Xe+^{124}Sn$ &  1.38  & PN & 0.54$\pm0.03$    & $\times$  \\
			\hline
            $^{129}Xe+^{nat}Sn$ &  1.38  &  & 0.55$\pm0.03$   & 0.53$\pm0.03$      \\
			\hline 
			$^{136}Xe+^{112}Sn$ &  1.38  & NP &0.50$\pm0.03$    & 0.54$\pm0.03$      \\
			\hline 
			$^{136}Xe+^{124}Sn$ &  1.5   & NN & 0.49$\pm0.03$    & 0.52$\pm0.03$      \\
			\hline
		\end{tabular}
	\end{center}
	\caption{Values of $E_{iso}^{cen}$ for different Xe+Sn reactions at 32 and 45 AMeV}
	\label{table1}
\end{table}

\begin{figure}[!htbp]
		\includegraphics[width=1\columnwidth]{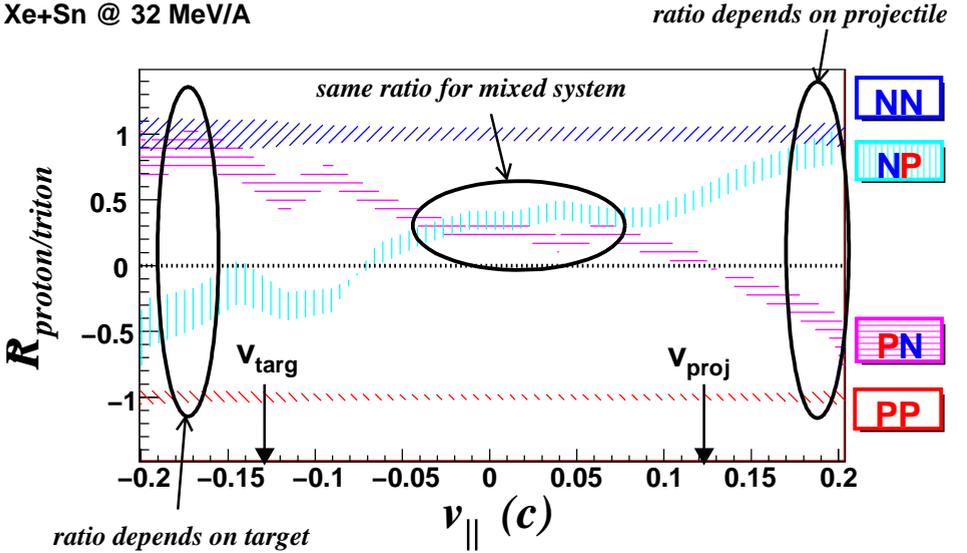}
	\caption{Ratio of proton/triton normalized as in Eq.~\ref{equRatio} along the beam axis for 
	Xe+Sn at 32 AMeV for different isospin reactions (for labels, see table~\ref{table1}).}
	\label{fig3}
\end{figure}

Mean values of $E_{iso}^{cen}$ for a given energy are similar within systematic errors in agreement with Liu's results.
To go further, we have been looking at the so-called isospin difusion by exploring the evolution of isospin distribution
along the beam axis using the double 
proton-triton~\cite{Ram00} ratio expressed as:

\begin{equation}
	\tilde{R}_{p/t} = \frac{2R_{p/t}-R_{p/t}^{NN}-R_{p/t}^{PP}}{R_{p/t}^{NN}-R_{p/t}^{PP}}
	\label{equRatio}
\end{equation}

where $R_{p/t}$ is the proton-triton ratio, $R_{p/t}^{NN}$ is the value of the ratio for the 
neutron-rich projectile neutron-rich target couple (i.e. $^{136}$Xe+$^{124}$Sn) and $R_{p/t}^{PP}$ is 
the value of the ratio for the neutron-poor projectile neutron-poor target couple (i.e. $^{124}$Xe+$^{112}$Sn). The ratio 
is shown in Figure~\ref{fig3} for different pairs of Xe+Sn isotopes at 32 AMeV.

For velocities larger than the projectile or smaller than the target, the  
proton-triton ratio is close to the projectile(target) isospin content suggesting a 
strong memory effect of the entrance channel even
for central collisions. As one moves towards the midvelocity region, the ratio smoothly evolves according 
to the isospin content of the compound system thus showing a mixing of the projectile and the target only 
for those particles emitted very close to mid-rapidity. Such results (together those shown in 
the previous section) are proofs of a sizeable 
transparency of nuclear matter in the intermediate energy range. 
The same behaviour is also observed for 45 AMeV.

\section{Summary}

Central collisions for a large variety of systems in the intermediate energy range have been 
studied using INDRA data. Charged particle kinetic energies have been used to build isotropy ratio 
in order to study nuclear stopping power. From 
10 to 40 AMeV, a strong decrease of the stopping power due to the decreasing role of the mean-field is observed while 
for energies larger than 40 AMeV, an increase is evidenced due to the 
increasing rate of in medium nucleon-nucleon collisions. However, our results show that 
tranparency is a key feature of nuclear collisions in the intermediate energy range. 
At 32 and 45 AMeV, stopping in Xe+Sn reactions with different isospin content do not show 
significative differences as observed in IQMD calculations. Last, the study of 
isospin diffusion along the beam axis confirms nuclear transparency but also reveals isospin mixing 
in the midrapidity region.

%%%%%%%%%%%%%%%%%%%%%%%%%%%%%%%%%%%%%%%%%%%%%%%%%%%%%%%%
% End of the paper
%
\end{document}